\documentclass[%
 aip,
 amsmath,amssymb,
 reprint,%
]{revtex4-1}

\usepackage{graphicx}
\usepackage{dcolumn}
\usepackage{bm}
\usepackage{enumerate}

\usepackage{setspace,color,url}
\begin{document}

\preprint{Submitted to The Journal of Chemical Physics}

\title{Labyrinthine water flow across multilayer graphene-based membranes: molecular dynamics versus continuum predictions}
%
\author{Hiroaki Yoshida}
\email{h-yoshida@mosk.tytlabs.co.jp}
\affiliation{LPS, UMR CNRS 8550, Ecole Normale Sup\'erieure, 24 rue
Lhomond, 75005 Paris, France}
\affiliation{Toyota Central R\&D Labs., Inc., Nagakute, Aichi 480-1192, Japan}
\author{Lyd\'eric Bocquet}
\email{lyderic.bocquet@ens.fr}
\affiliation{LPS, UMR CNRS 8550, Ecole Normale Sup\'erieure, 24 rue
Lhomond, 75005 Paris, France}
\date{\today}
%
\begin{abstract}

In this paper we investigate the hydrodynamic permeance of water through graphene-based membranes, inspired by recent
experimental findings on graphene-oxide membranes. We consider the flow across multiple graphene
 layers having nanoslits in a staggered alignment, 
 with an inter-layer distance ranging from sub-nanometer to a few nanometers.
 We compare results for the permeability obtained by means of molecular dynamics simulations to continuum
 predictions obtained by using the lattice Boltzmann calculations and hydrodynamic modelization.
This highlights that, in spite of extreme confinement, the permeability across the graphene-based membrane is quantitatively predicted
on the basis of a continuum expression, 
taking properly into account entrance and slippage effects
of the confined water flow. Our predictions refute the breakdown of hydrodynamics at small scales in these membrane systems. They 
constitute a benchmark to which we compare published experimental data.

\end{abstract}
\maketitle

%
%
\section{\label{sec_intro}Introduction}

Recent progress in chemical modification and conversion technology 
concerning graphene sheets has opened up the possibility of producing a bulk 
carbon material having a molecular-scale porous structure
with  well-controlled pores and inter-layer distances,
as represented by the graphene oxide (GO) membrane.~\cite{DSZ+2007,NFC+2012,JAY+2015,YCP2016}
In contrast to the graphite, 
in which the graphene sheets are held together by
van der Waals forces with the distance around $3.4$\,{\AA}
and there is no space for fluid molecules, 
the inter-layer distance in GO membrane is maintained typically 
at $\sim1$\,nm, {\it i.e.}, a few times larger than fluid molecules,
and it works as a membrane for fluids flowing through the gap between
the layers. 
The inter-layer distance 
may be systematically tuned in the range of sub-nanometers up to $10$\,nm.~\cite{YCW+2013}
Such graphene-based materials have attracted significant attentions as
multi-functional membranes that exhibit
peculiar transport phenomena relevant to 
nano-scale flows.~\cite{QZY+2011,NWJ+2012,SZW+2013,CL2013,HSW+2013,HYP2014,PJ2014,AAW+2015,HZ2015,CJG+2016,CET2016,KH2016}
In Ref.~\onlinecite{NWJ+2012}, a GO film fabricated to have pores (or slits) 
was shown to allow high-speed water flow across the film,
whereas it was almost completely impermeable to any other liquids or gases.
The GO film in Ref.~\onlinecite{HSW+2013},
which was designed to have channels of $3-5$\,nm in width,
was shown to permeate water very efficiently.
More recently, ion transports through nanoslits 
in stacking multiple graphene sheets
have been examined and 
the phenomena specific to the complex geometries have been reported.~\cite{CJG+2016}

Since the inter-layer distance ranges
from a few to tens of the size of fluid molecules,
the transport phenomena unique to multi-layered graphene membranes
observed experimentally
are often attributed to atomic-scale effects
that can not be addressed in the continuum theory. 
However,
a systematic investigation of the flow across such complex porous structure,
which lies at the edge between 
the atomic scale and the continuum framework, is still lacking.
In particular,  it is still difficult to clarify whether the 
peculiar transport phenomena observed experimentally
are indeed dominated by the breakdown of hydrodynamics, the specific geometrical complexity of
a small-scale structure, or caused by other effects 
such as the surface chemistry of the modified graphene.~\cite{BKS2013,WPX2014,BXW+2016}

In the present study, 
we investigate the 
water flow across a multi-layered graphene
with arrays of staggered nanoslits
using the molecular dynamic (MD) simulation,
and develop a 
corresponding continuum model for comparison, in order
to clarify the applicability of classical hydrodynamics and to benchmark its predictions.
Focusing on the influence of the geometry,
we assume that the layers consist of pure graphene sheets
without chemical modification,
but the width of the nanoslits and the inter-layer distance
are controllable from several angstroms to a few nanometers,
mimicking the porous structure of GO
membranes.~\cite{NWJ+2012,XNZ+2015,HM2013,BXW+2016,ASM+2016}
Here, we measure the water flux across the membrane 
and make systematic comparison with the permeance predicted by the
developed continuum model.
Good agreement is obtained if the model parameters are chosen
appropriately,
the values of which are discussed by analyzing basic problems 
such as an independent nanoslit and a flow through two parallel graphene sheets.

%
%
\section{\label{sec_problem}Geometrical set-up and 
MD simulations}

\begin{figure}[t]
\begin{center}
\includegraphics[scale=0.72]{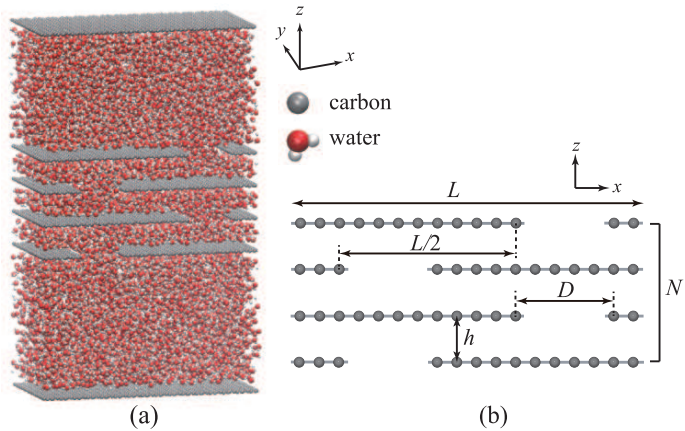} 
\caption{
(a) A snapshot of the system.
(b) Geometrical parameters characterizing the multi-layered graphene membrane.
}
\label{fig-geometry}
\end{center}
\end{figure}  

We consider a physical model 
of multi-layered graphene membrane as depicted in
Fig.~\ref{fig-geometry},
where graphene sheets having nanoslits of width $D$ are laminated in the $z$ direction.
The nanoslits are arranged in a staggered fashion
such that the displacement in the $x$ direction is $L/2$, 
and the common inter-layer distance is $h$.
The membrane is sandwiched by two water reservoirs, and 
each of the two ends is closed by a graphene sheet with no slit.
The graphene sheet at the ends plays the role of a piston
controlling the pressure in the reservoir.
The periodic boundary condition is assumed in the $x$ and $y$ directions.
The system is considered as a pseudo two-dimensional 
problem (Fig.~\ref{fig-geometry}(b)), as treated in Refs.~\onlinecite{NWJ+2012,JCW+2014}.
The porous structure of the 
membrane is 
characterized by the four geometrical parameters, namely, 
the periodicity $L$ in the $x$ direction,
the width $D$ of the nanoslit,
the inter-layer distance $h$,
and the number of graphene layers $N$.
Here, it is emphasized that $D$ and $h$ are defined as distances
between the centers of carbon atoms (in contrast with the continuum
model discussed in Sec.~\ref{sec_model}).

In the MD simulation, 
the interaction potential employed for water molecules is the TIP4P model.~\cite{JCM+1983}
The water-graphene interaction potential is
determined by 
the Lorentz--Berthelot mixing rule,~\cite{AT1989,HM2006}
employing the Lenard--Jones (LJ) parameters
of AMBER96 for carbon atoms.~\cite{CCB+1995}
The contact angle of a water droplet on a pure graphene sheet is
$66^{\circ}$, which we evaluated using the method described in 
Ref.~\onlinecite{WWJ+2003},
and thus the surface of the graphene in the present study is hydrophilic.

\begin{figure}[t]
\begin{center}
\includegraphics[scale=0.9]{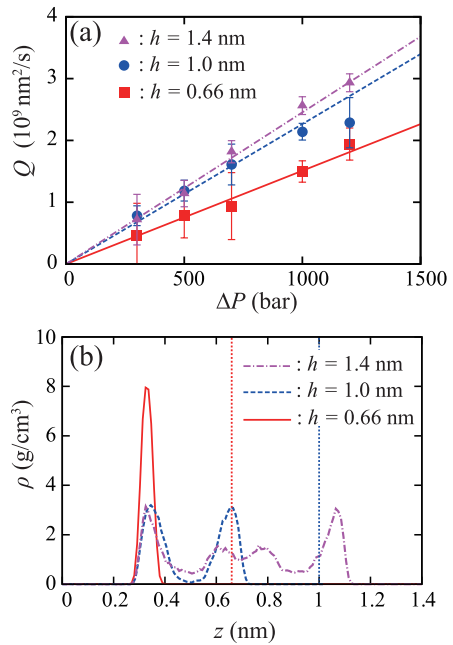} 
\caption{(a) Volumetric flux $Q$ per unit length in the $y$ direction versus
 applied pressure difference $\Delta P$.
The geometrical parameters are $D=0.99$\,nm, $L=6.82$\,nm, and $N=2$.
The cases of $h=0.66$, $1.0$, and $1.4$\,nm are shown in the figure.
The error bar indicates the standard deviation for the data measured at every
$1$\,ps (see also the main text). The linear fit for each $h$ is also
 indicated by the lines.
{\color{black}
(b) The cross-sectional density distribution of water between two
 graphene layers.
The origin of the $z$ coordinate is the position of one graphene layer,
and the positions of the other graphene layer for
 the cases of $h=0.66$ and $1.0$\,nm are indicated by the dotted line.
}
}
\label{fig-vs_p}
\end{center}
\end{figure}  

The MD simulations are implemented using the open-source code
LAMMPS.~\cite{LAMMPS}
The number of molecules and the size of the simulation box are 
fixed during each simulation,
while the temperature is maintained at $300$\,K using the Nos\'e--Hoover
thermostat (NVT ensemble).
The time integration is carried out with the time step $1$\,fs,
and SHAKE algorithm is employed to maintain the water molecules as rigid.~\cite{RCB1977}
The LJ interactions are treated using the standard method with spherical
cutoff of $9.8$\,{\AA}, while the long-range Coulomb interactions are treated
using the particle-particle particle-mesh (PPPM) method.
The non-periodicity in the $z$ direction
is dealt with by applying the periodic boundary condition
with empty spaces outside the pistons, 
and the artifacts from the image charges due to
periodic conditions in the $z$ direction are removed by 
using the method in Ref.~\onlinecite{YB1999}.

The pressure in the reservoirs is controlled by tuning
the force acting on the pistons.
More precisely, the atoms of the pistons are constrained
such that they move only in the $z$ direction, and 
the force on all atoms is tuned so that the  force per unit
surface corresponds to the desired pressure.

The water permeance is evaluated by measuring 
the fluxes induced by the pressure difference $\Delta P$ between the reservoirs.
Before running each simulation with the pressure difference,
the system is equilibrated for $0.5$\,ns
with maintaining the pressures of both reservoirs at $1$\,bar ($10^5$\,Pa).
The size of the simulation box is $1.97\sim 3.94$\,nm in the $y$ direction,
and the initial height of the reservoirs in the $z$ direction is more than $3.5$\,nm.
In Fig.~\ref{fig-geometry}(a), the length in the $y$ direction is
$3.94$\,nm, and the geometrical parameters are $L=6.82$\,nm, $D=1.42$\,nm,
$h=1$\,nm, and $N=4$. The number of molecules contained in this case
is $8640$.

Figure~\ref{fig-vs_p}(a) plots the volumetric flux $Q$
as a function of $\Delta P$ in the cases of  
$h=0.66$, $1.0$, and $1.4$\,nm with $D=0.99$\,nm, $L=6.82$\,nm, and $N=2$.
The volumetric flux is obtained from the linear fit of the motion of
piston $z_p(t)$.
The time-series values of $z_p(t)$ averaged over $1$\,ps at every $1$\,ps are used for the linear fit,
and the standard deviation is 
shown by the error bar in the figure. 
Although the error bar is large for the small values of $\Delta P$, 
the measured flux is found to increase linearly with $\Delta P$, and
the water permeance ${\cal L}_{\rm hyd}=Q/\Delta P$ is obtained independent of 
$\Delta P$ in the range considered here.

{\color{black}
The water density between two graphene sheets 
during the production runs for Fig.~\ref{fig-vs_p}(a) is shown in
Fig.~\ref{fig-vs_p}(b). 
The density distribution
along the $z$ axis at the midpoint of the two nanoslits
 staggered in the $x$ direction is plotted,
for the case of $\Delta P=1000$\,bar.
The excluded volume near the carbon atoms in the graphene
is clearly observed, which should be
properly taken into account in the 
continuum model. The consistency of this density profile
with the model parameter describing
this excluded volume will be shown in Sec.~\ref{subsec_parameter}
in the course of determination of the parameters contained in the continuum model.
}

\begin{figure}[t]
\begin{center}
\includegraphics[scale=1.1]{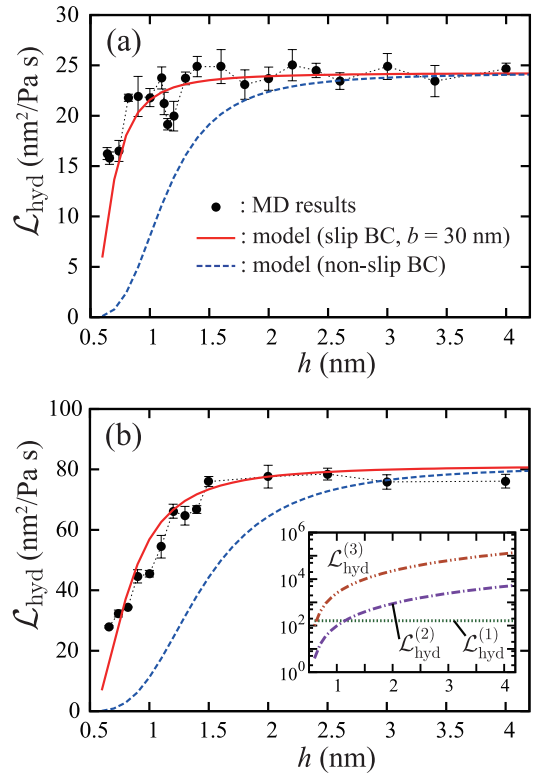} 
\caption{
Two-dimensional hydrodynamic permeance ${\cal L}_{\rm hyd}$ versus the inter-layer distance $h$ obtained by means of the MD simulations. The solid and dashed lines are the prediction of the continuum model in Eq.~\eqref{s3-gm}, 
with respectively a slip length of $b=30$\,nm and no-slip boundary conditions, see Sec.~\ref{sec_model}.
The geometrical parameters are $L=6.82$\,nm, $N=2$ 
and $D=0.99$\,nm in panel (a) or $D=1.41$\,nm in panel (b).
The error bar indicates the standard deviation for the data obtained
for different values of the pressure differences $\Delta P$.
{\color{black}
The inset of panel (b) shows the contributions
of decomposed permeances defined in Sec.~\ref{sec_model},
in the case of $b=30$\,nm.
}
}
\label{fig-md}
\end{center}
\end{figure}

In Fig.~\ref{fig-md}, we show the MD results for the permeance as a function of the
inter-layer distance.
Since the flux $Q$  increases linearly with $\Delta P$
as shown in Fig.~\ref{fig-vs_p}, 
the permeance is evaluated for at least four values of 
$\Delta P$ in the range $\Delta P\le 1200$\,bar and the averaged value is plotted.
The standard deviation for different values of $\Delta P$
is shown by the error bars in the figure.
The geometrical parameters are $L=6.82$\,nm, $N=2$, and 
$D=0.99$\,nm in Fig.~\ref{fig-md}(a) or $D=1.42$\,nm in Fig.~\ref{fig-md}(b).
Since the diameter of a water molecule is about $3$\,{\AA},
the width of the nanoslit is twice as large as a water molecule in Fig.~\ref{fig-md}(a),
and three times in Fig.~\ref{fig-md}(b), taking into account the
excluded volume near the carbon atoms {\color{black}(cf. Fig.~\ref{fig-vs_p}(b))}.
The model equations plotted by the lines will be derived 
in the following section, and the MD results in comparison with the model
predictions will be discussed in Sec.~\ref{sec_result}. 

%
%

%
%
\section{\label{sec_model}Continuum model of hydrodynamic permeance}

\begin{figure}[t]
\begin{center}
\includegraphics[scale=0.75]{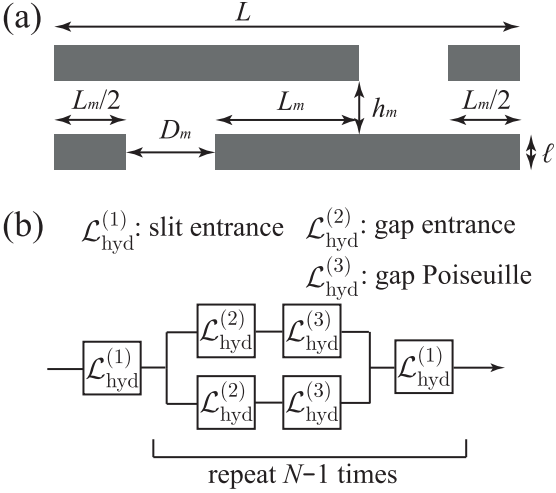} 
\caption{(a) Coarse-grained geometry of the membrane. (b) Continuum
 model of water permeance.
}
\label{fig-conductance}
\end{center}
\end{figure}  

In this section, we develop a model based on continuum hydrodynamics 
in order to predict the flow through the membrane
in Fig.~\ref{fig-geometry}.
To this end, we define a two-dimensional channel
depicted in Fig.~\ref{fig-conductance}(a),
as a coarse-grained model for the
original porous structure of carbon atoms.
The definition of the width of the slit $D_m$ and the inter-layer distance $h_m$
differs from the molecular parameters $D$ and $h$ defined in Fig.~\ref{fig-geometry}(b)
in terms of the distance between carbon atoms,
due to the exclusion of water molecules close to the graphene surfaces {\color{black}(see Fig.~\ref{fig-vs_p}(b))}. 
At this stage, we introduce exclusion distances for the slit width and inter-layer
distance, defined as $\delta D=D-D_m$ and $\delta h=h-h_m$,
respectively. 

The model is built by decomposing the flow from
the entrance to the exit of the membrane into three parts.
First, the permeance describing the flow resistance at the entrance
of a nanoslit is written as:~\cite{Hasimoto1958}
\begin{equation}
{\cal L}_{\rm hyd}^{(1)}=\dfrac{\pi D_m^2}{32\eta},
\label{s3-g1}
\end{equation}
where $\eta$ is the viscosity of water.
This is the two-dimensional version of the Sampson formula 
for the flow through a single circular pore in an infinitely thin wall.~\cite{Sampson1891}
Note that this is a hydrodynamic permeance per
unit length in the $y$ direction (with unit  m$^2/$Pa\,s). 

The effect at the entrance into the gap $h_m$ between the layers 
is described by essentially the same formula as Eq.~\eqref{s3-g1}. 
A slight modification is necessary, however, because there is only one
edge at this entrance, and the other side is in contact with the plane surface.
We consider this situation as the half of the entrance of width $2h_m$,
which results in:
\begin{equation}
{\cal L}_{\rm hyd}^{(2)}=\dfrac{1}{2}\dfrac{\pi (2h_m)^2}{32\eta}.
\label{s3-g3}
\end{equation}

Finally, the permeance of the flow through the gap $h_m$ of length
$L_m$ is given by 
{\color{black}
the formula for the plane Poiseuille flow
with Navier's slip boundary condition:}
\begin{equation}
{\cal L}_{\rm hyd}^{(3)}=\dfrac{h_m^3}{12\eta L_m}+\dfrac{bh_m^2}{2\eta L_m},
\label{s3-g4}
\end{equation}
{\color{black}
where the second term on the right-hand side arises from
the slip boundary condition with $b$ being the slip length.
}

The hydrodynamic permeance of the whole membrane is obtained by combining
${\cal L}_{\rm hyd}^{(1)}\sim {\cal L}_{\rm hyd}^{(3)}$ as in Fig.~\ref{fig-conductance}(b).
Since the permeances  in series are combined through the harmonic mean
while those in parallel are simply added, the complete model is
expressed as:
{\color{black}
\begin{equation}
{\cal L}_{\rm hyd}=\left[\dfrac{N}{{\cal L}_{\rm hyd}^{(1)}}+(N-1)\left(\dfrac{{\cal L}_{\rm hyd}^{(2)}+{\cal L}_{\rm hyd}^{(3)}}{2{\cal L}_{\rm hyd}^{(2)}{\cal L}_{\rm hyd}^{(3)}}\right)\right]^{-1}.
\label{s3-gm}
\end{equation}
}

In order to verify the accuracy of Eq.~\eqref{s3-gm} at this stage {\it within the
continuum description}, 
we carry out a direct numerical analysis of the Navier--Stokes
equations for the geometry in Fig.~\ref{fig-conductance}(a).
We employ the lattice Boltzmann method (LBM)~\cite{CD1998,S2001} as the 
numerical method,
the detailed algorithm of which is described in Ref.~\onlinecite{YH2014}.
The no-slip boundary condition is implemented using the standard
halfway bounce-back rule, and the perfect-slip condition is realized 
with the specular reflection. 
At a boundary sufficiently far form the membrane in the $z$ direction, 
a pressure difference of $\Delta P=1$\,bar is imposed using the method
in Ref.~\onlinecite{ZH1997}. The hydrodynamic permeance ${\cal L}_{\rm hyd}$ is then evaluated 
by measuring the flux $Q$ in the $z$ direction.

\begin{figure}[t]
\begin{center}
\includegraphics[scale=1.1]{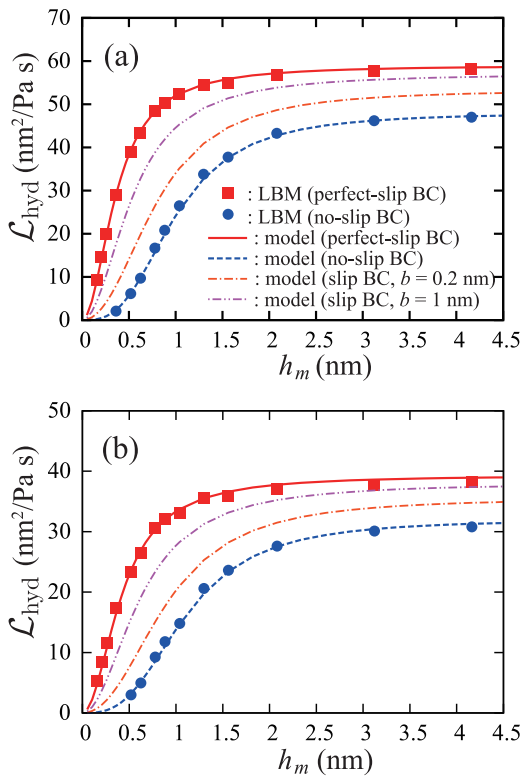} 
\caption{
Two-dimensional hydrodynamic permeance ${\cal L}_{\rm hyd}$ versus the inter-layer distance $h_m$ obtained by means of the LBM, in comparison with the continuum model
in Eq.~\eqref{s3-gm} (Eq.~\eqref{s3-gm2} for the perfect slip boundary
 condition.)
The geometrical parameters are $D_m=1$\,nm, $L_m=3.7$\,nm, $\ell=0.31$\,nm,
 and $N=2$ in panel (a) or $N=3$ in panel (b).
}
\label{fig-lbm}
\end{center}
\end{figure}

The hydrodynamic permeance ${\cal L}_{\rm hyd}$ predicted by the LBM is plotted as a function of $h_m$ 
in Fig.~\ref{fig-lbm}.
The geometrical parameters used in the LBM are
$D_m=1$\,nm, $L_m=3.7$\,nm, $\ell=0.31$\,nm and $N=2$
in Fig.~\ref{fig-lbm}(a) or $N=3$ in Fig.~\ref{fig-lbm}(b).
The results of Eq.~\eqref{s3-gm} with $b=0$ (no-slip), $0.2$ and $b=1$\,nm are also shown
in the figure.
The permeance of the perfect-slip case shown in the 
figure is obtained by taking the limit
of $b\to \infty$ in Eq.~\eqref{s3-gm}:
\begin{equation}
{\cal L}_{\rm hyd}=\left(\dfrac{N}{{\cal L}_{\rm hyd}^{(1)}}+\dfrac{N-1}{2{\cal L}_{\rm hyd}^{(2)}}\right)^{-1}.
\label{s3-gm2}
\end{equation}
In the model equations, the same values of the geometrical parameters as
those in the LBM are used.
{\color{black}
Note furthermore that for completeness, the dissipation across the  
thickness $\ell$ of the slit could be accounted for. A crude approximation
consists in using a Poiseuille like dissipation, with a permeance given as
$D_m^3(1+6 b/D_m)/(12\eta \ell)$. This is actually a small correction as compared to ${\cal L}_{\rm hyd}$
in Eq.~(\ref{s3-gm}), but using a value $\ell=0.2$\,nm allows to reach perfect agreement with
numerical LBM results, as shown in Fig.~\ref{fig-lbm}.
As a further remark, we quote that the perfect-slip case ($b\to \infty$, {\it i.e.} $b\gg D_m,\, h_m$) does not contain $\ell$, so that no parameter is tuned.}
It is clear form the figure that the model in Eqs.~\eqref{s3-gm}
reproduces the LBM results very
accurately. As a conclusion, the prediction Eq.~\eqref{s3-gm} is a quantitative prediction 
for the permeance within the continuum framework.

%

%
%
\section{\label{sec_result}Comparison of MD results with hydrodynamic predictions}

We now gather the various results and compare the water permeance obtained using the MD 
simulations in Sec.~\ref{sec_problem} with the model prediction
in the previous section.

\subsection{\label{subsec_parameter}Flow parameters}

The comparison between the MD simulations and hydrodynamic calculation
requires to define several quantities: the viscosity, the slip length $b$, and the corrections $\delta D$ and $\delta h$, which determine the effective
lengths $D_m$ and $h_m$, respectively. The value of the viscosity used in the hydrodynamic
model is taken from Ref.~\onlinecite{GA2010}. It is evaluated for the interaction potential
used in the present study.
In order to determine unambiguously the values of 
$b$,  $\delta D$ and $\delta h$, we consider alternative geometries: (i) the flow through a slit across
a single layer graphene, 
as well as (ii) a slab geometry with water confined between two graphene walls to determine the slip length.

\begin{figure}[t]
\begin{center}
\includegraphics[scale=0.8]{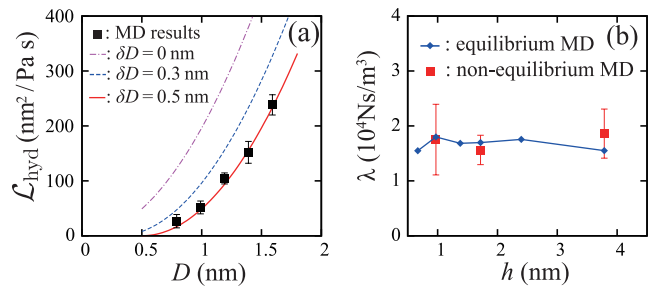} 
\caption{
(a) Two-dimensional hydrodynamic permeance of an independent slit obtained using the
 MD simulation measured at $\Delta P=1000$\,bar, in comparison with the continuum model
in Eq.~\eqref{s3-g1} with different values of $\delta D$. See the caption of
 Fig.~\ref{fig-vs_p} for the meaning of the error bar. 
(b) Friction coefficient $\lambda$ for the slab geometry of the gap $h$.
The results of the equilibrium and non-equilibrium MD simulations are
 shown. The error bar indicates the standard deviation for the data obtained
with different values of the applied body force.}
\label{fig-ps}
\end{center}
\end{figure}

First, in Fig.~\ref{fig-ps}(a), 
the permeance of an independent nanoslit across a single layer of graphene is plotted versus
the slit width.  
These data are compared to the 
corresponding continuum model ${\cal L}_{\rm hyd}^{(1)}$ given in Eq.~\eqref{s3-g1}, in order to estimate 
the effective hydrodynamic width $D_m$. 
The results of Eq.~\eqref{s3-g1} for a few values of $\delta D$ are shown.
If one sets $D_m=D$ (or $\delta D=0$), the permeance is overestimated, as
expected considering the excluded volume around the carbon atoms.
Clearly, the choice of $\delta D=0.5$\,nm does  approximate well
the excluded volume and yields a precise prediction for the permeance.
The gap correction $\delta h$, which also accounts for
exclusion effects, is  expected to be quantitatively similar to 
$\delta D$, {\it i.e.} 
$\delta h \approx 0.5$\,nm.
This was actually checked in the comparison between the MD results for
the flux in the slab geometry described below
and the hydrodynamic model ${\cal L}_{\rm hyd}^{(3)}$ in Eq.~\eqref{s3-g4} with different values of $\delta h$
(not shown). As a result, the choice of $\delta h=0.5$\,nm (or $h_m=h-0.5$\,nm) 
is found to give a good agreement. 
{\color{black}
We note that the values for the exclusion volume 
determined here in terms of the measurement of the flux,
{\it i.e.} $\delta D=\delta h=0.5$\,nm, are consistent with the
density profile shown in Fig.~\ref{fig-vs_p}(b),
and a relevant discussion is also found in Ref.~\onlinecite{GJY+2014}.
}
In the following, we set accordingly $\delta D=\delta h=0.5$\,nm (or
$D_m=D-0.5$\,nm, $h_m=h-0.5$\,nm) 
to compare with MD data in Fig.~\ref{fig-md}.

In order to estimate the slip length $b$,
we next consider a water slab confined between two 
parallel graphene sheets with no slit. We
examine the friction coefficient $\lambda$ between the water and
the graphene walls by means of the methods described in Ref.~\onlinecite{FSJ+2012}.
The slip length is then evaluated from the friction coefficient via the
relation $b=\eta/\lambda$.
The friction coefficient is obtained using two different methods.
First, we measure the fluctuation of the friction force $F$ (force acting in the
lateral direction to the water form the graphene) at an equilibrium state without
external force, in order to obtain the friction coefficient through the Green--Kubo formula.~\cite{BB1994,BB2007}
Second we measure the average slip velocity $v$ of  water
during a non-equilibrium MD simulation with a constant force acting on each water molecule in the 
direction parallel to the graphene sheet.
The friction coefficient is then directly evaluated using the relation
$\lambda=-F/Av$ ($A$ is the area of the sheet).
Figure~\ref{fig-ps}(b) shows the obtained friction coefficient as a
function of the gap between the two sheets.
As in Ref.~\onlinecite{FSJ+2012},  the friction coefficient is measured to be
  independent of the gap.
The resulting slip length is $b=30$\,nm, which is also consistent with
the previous results,~\cite{TM2008,FSJ+2012,KTH+2012}
and this value is used in Fig.~\ref{fig-md}.

\begin{table}[t]
\vspace*{0cm}
\centering
\caption{
Model parameters.
}
\label{table01}
\begin{tabular}{ll}
\hline
\hline
 slip length $b$ & $30$\,nm\\
 slit width correction $\delta D$ & $0.5$\,nm \\
 inter-layer distance correction $\delta h$ & $0.5$\,nm \\
\hline
\hline
\end{tabular}
\end{table}

\begin{figure}[t]
\begin{center}
\includegraphics[scale=0.8]{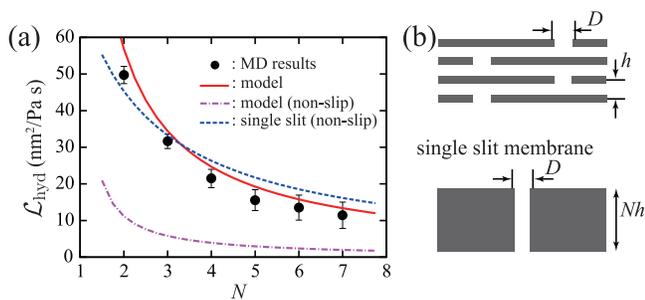} 
\caption{
(a) Two-dimensional hydrodynamic permeance ${\cal L}_{\rm hyd}$ versus number of graphene layers $N$.
The symbol indicates the MD results evaluated at $\Delta P=1000$\,bar,
and the lines indicate the model predictions. 
The dashed blue line is the predicted permeance of an single-slit membrane
depicted in panel (b).
The geometrical parameters are $L=6.82$\,nm, $h=1$\,nm, 
and $D=1.41$\,nm.
See the caption of Fig.~\ref{fig-vs_p} for the
 meaning of the error bar.
}
\label{fig-nlayers}
\end{center}
\end{figure}

\begin{figure}[t]
\begin{center}
\includegraphics[scale=0.8]{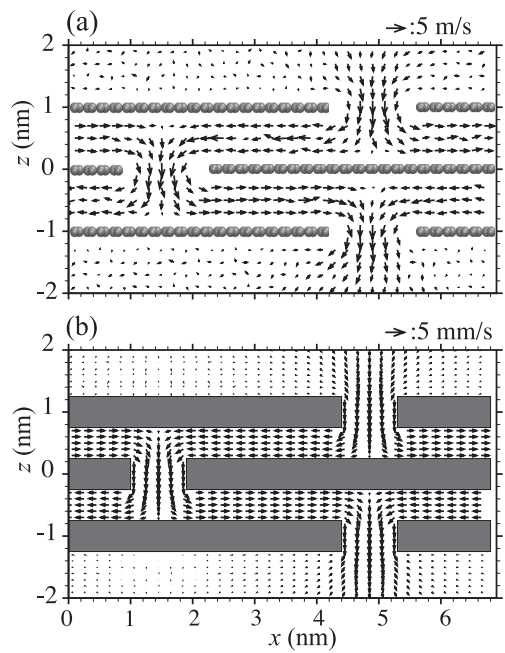} 
\caption{
\color{black}
Flow velocity profiles in the $x$-$z$ plane. (a) MD result for the 
case of $L=6.82$\,nm, $h=1$\,nm, $D=1.41$\,nm, and $N=3$ obtained
 applying the pressure difference $\Delta P=1000$\,bar, and (b) the
 corresponding profile obtained using the LBM with the perfect-slip
 boundary condition applying $\Delta P=1$\,bar.
 The scale of the vector is shown above each panel.
}
\label{fig-prof}
\end{center}
\end{figure}

Table~\ref{table01} lists the 
model parameters for Eq.~\eqref{s3-gm} determined 
from the discussion above, 
and Fig.~\ref{fig-nlayers}
shows the comparison of the MD results for $N\ge 2$ with the
model prediction using these parameters.
The geometrical parameters are $L=6.82$\,nm, $h=1$\,nm, and $D=1.42$\,nm.
The MD results are the permeance evaluated at $\Delta P=1000$\,bar.
Even for the very complex geometry with many entrances and gaps up to $N=7$,
the permeance is well predicted by the model.
{\color{black}
A typical profile of the flow velocity in the case of $N=3$ is also shown in
Fig.~\ref{fig-prof}, in comparison with the corresponding profile obtained using
the LBM.
}
As mentioned above, since the slip length is very large and 
the model is almost identical to the perfect-slip case,
the decreasing of the permeance as $N$
is caused by the increase of the number of entrances,
rather than the increase of the length of the flow path.


\subsection{\label{subsec_compare}Multi-layered  graphene membrane: MD versus hydrodynamics}

We can now discuss the MD results for the permeance, as shown in Fig.~\ref{fig-md}, 
in light of hydrodynamic predictions in Eq.~\eqref{s3-gm}. 

A first result is that  the model with the no-slip boundary condition ($b=0$) greatly
underestimates the permeance for $h\le 2$\,nm. Now the hydrodynamic
model with $b=30$\,nm for the slip length, as obtained above, gives a good
agreement with the MD results for various inter-layer distance and for the two typical
conditions in Fig.~\ref{fig-md}.
Altogether this comparison confirms that
the continuum framework provides a quantitative prediction for the permeance down to
sub-nanometer gaps between the layers, if the model parameters are 
appropriately chosen.

More into the details of the results, one observes two regimes for the permeance in Fig.~\ref{fig-md}.
One is the range $h>D$ where the flow is dominantly limited by the resistance
at the entrance of the nanoslits (${\cal L}_{\rm hyd}^{(1)}$).  The dependence on $h$ is thus
relatively {\color{black}weak} in this regime. On the other hand,
the resistance at the entrance of the gap (${\cal L}_{\rm hyd}^{(2)}$) starts to limit
the flow at $h\sim D$ and the permeance rapidly decreases as
decreasing $h$ in the range $h<D$.  
{\color{black}
The quantitative explanation of this scenario is given by the plot of the decomposed
permeances in the inset of Fig.~\ref{fig-md}(b), and 
consequently 
}
the hydrodynamic model in Eq.~\eqref{s3-gm} accurately
captures this two-regime behavior.
{\color{black}
The validity of this hydrodynamic model in atomic scale is further
supported by the very recent MD results for a similar geometry in Ref.~\onlinecite{MJM+2016} 
(published after the submission of the present work,)
where a detailed molecular analysis of the hydrogen-bonding shows a sufficient mixing
of water even for a small inter-layer distance down to $0.6$\,nm, thus promoting 
hydrodynamic behavior.
}

Finally, in Fig.~\ref{fig-nlayers}, the MD results for the multi-layer graphene membrane are compared to the permeance of an artificial 
single-slit, solid, membrane with thickness $Nh$. 
We define the permeance of the single-slit
geometry as ${\cal L}_{\rm hyd}=(1/{\cal L}_{\rm hyd}^{(1)}+1/{\cal
L}_{\rm hyd}^{(3)})^{-1}$ 
{\color{black} 
with $L_m$ and $h_m$ in Eq.~\eqref{s3-g4} replaced by $Nh$ and $D_m$, respectively, 
}
and the boundary condition inside the slit is assumed to be that of {\it no-slip}
condition ($b=0$).
The permeance of this single-slit no-slip membrane is found to be comparable
to that of the multi-layered graphene membrane. 
This means that, in the nano-structured membrane, the slip effect
becomes significant and compensates the reduction of the permeance due
to the labyrinthine complexity of the geometry.
The significance of this effect is clear if one compares
with the model prediction for the multi-layer membrane with the no-slip condition, as shown in Fig.~\ref{fig-nlayers} (bottom curve).
Also, if one assumes $b=30$\,nm for the single-slit case,
the permeance is far larger than that in Fig.~\ref{fig-nlayers}(a)
because  the main resistance is that of the single slit entrance.

\begin{table}[b]
 \begin{center}
 \hspace*{-0mm}
 \begin{minipage}{0.37\textheight}
\small
\centering
\caption{Comparison with experimental data.}
\label{table02}
\vspace*{2mm}
 \begin{tabular}{ccccccc}
\hline
\hline
 & ${\cal L}_{\rm A}^{\rm (experiment)}$ & ${\cal L}_{\rm A}^{\rm (model)}$ & $h\,^{\rm a)}$ &
 $D\,^{\rm b)}$ & $L\,^{\rm b)}$
 & $N\,^{\rm a)}$ \rule[0mm]{0mm}{4mm}\\
 & {\scriptsize(nL$/$m$^2$Pa\,s)} & {\scriptsize(nL$/$m$^2$Pa\,s)} & {\scriptsize(nm)} & {\scriptsize(nm)} & {\scriptsize(nm) }\\
\hline
Xia {\it et al.}\cite{XNZ+2015}
 & $75.3$ & $74.2$ & $1.38$ &  $10$ & $190$ & $14$
\rule[0mm]{0mm}{4mm}\\
$\uparrow$
 & $30.6$ & $23.3$ & $0.99$ &  $10$ & $190$ & $14$
\rule[0mm]{0mm}{4mm}\\ 
Hu\,\&\,Mi\cite{HM2013}
 & $62.9$ & $62.3$ & $1.75$ &  $10$ & $315$\,$^{\rm c)}$ & $15$
\rule[0mm]{0mm}{4mm}\\
Nair {\it et al.}\cite{NWJ+2012}
 & $10^4$ & $0.19$ & $1.0$ &  $10$ & $1000$\,$^{\rm c)}$ & $100$
\rule[0mm]{0mm}{4mm}\\
\hline
\hline
\end{tabular}
\vspace{-2mm}
\begin{flushleft}
$^{\rm a)}$ taken from the references. 
$^{\rm b)}$ estimated.
$^{\rm c)}$ estimated within the range provided in the references.
\end{flushleft}
 \end{minipage}
 \end{center}
\end{table}
%

%
%
%
\section{Concluding remarks}

In the present study, we investigated
the hydrodynamic permeance of the water flow past the geometrically 
complex graphene-based membrane,
by means of a combined analysis 
of the atomic-scale MD simulation and the
continuum hydrodynamics framework.
To construct the continuum model,
we first defined the coarse-grained 
geometry approximating the original
configuration of the atomic-scale
graphene walls. 
Then the derived simplified model was proven to be sufficiently accurate
within the continuum description, by comparing with
the results of the direct numerical simulation using the lattice
Boltzmann method.
The parameters appearing in the model from the coarse-graining of the
geometry, {\it i.e.}, the slip length $b$ and the corrections to
the slit width $\delta D$ and to the gap $\delta h$, were identified as in Table~\ref{table01},
by means of the MD simulations of the decomposed basic problems.
With these values of the parameters, the model was shown to predict the full MD results as shown in
Figs.~\ref{fig-md} and \ref{fig-nlayers}.

An important consequence in the present study 
is that the continuum description
is still valid for explaining
the small-scale flows down to sub-nanometers,
if the parameters are
carefully chosen.
This is a benchmark result which is essential to
examine critically the experimental results 
reported for graphene-based membranes.
An example is shown 
in Table~\ref{table02}, where the experimental results
reported in Refs.~\onlinecite{XNZ+2015,HM2013,NWJ+2012} are compared to
the present model in Eq.~\eqref{s3-gm}. 
In the table, the water permeance
per unit area of the membranes ${\cal L}_{\rm A}$ is listed,
which is related to ${\cal L}_{\rm hyd}$ via ${\cal L}_{\rm A}={\cal L}_{\rm hyd}/L$.
The present model exhibits good agreements
with the experimental results in Refs.~\onlinecite{XNZ+2015,HM2013},
if the values of the geometrical parameters $D$ and $L$
are estimated within reasonable ranges, 
for which the precise values are unavailable in the references.
(Since the geometries in the experiments are not strictly identical
to the setup considered in the present study, the tuned
values are regarded as the effective values for
the experimental membranes that include random configurations.
Note that in the regime of $D\gg h$, the permeance is barely sensitive to $D$.)

On the other hand, the present model strongly underestimates the result
of Ref.~\onlinecite{NWJ+2012}, suggesting that other effects may contribute to the giant
permeance measured in Ref.~\onlinecite{NWJ+2012}. 
A possible reason lies in the driving force
used to measure the permeability. 
While in Refs.~\onlinecite{XNZ+2015,HM2013}
it is obtained with an imposed pressure drop, in Ref.~\onlinecite{NWJ+2012}
water evaporation is used
to measure the permeability. In this case, a very large capillary contribution to the disjoining
pressure due to the nanometric inter-layer distance~\cite{GYB+2016}
may add up to the imposed pressure drop and increase
accordingly the driving force. This supplementary capillary pressure
would lead to a strong flow in spite of a small imposed pressure
drop, thus affecting the extracted value of the permeance. 
Note that reversely the independent knowledge of the permeance in
this system would allow to get much insights into the capillary pressure
and disjoining effects at small scales.~\cite{GYB+2016}
This suggests further experimental work along these lines.

One of extensions of the present study would be 
investigating flow properties of fluids other than water,
and exploring the performance as filtration and separation membranes. 
A validation of the model prediction, by comparing 
with the permeance observed experimentally under a well-defined situation, 
would also be an important topic.


\begin{acknowledgments}
This work was granted access to the HPC resources of MesoPSL financed
by the Region Ile de France and the project Equip@Meso (reference
ANR-10-EQPX-29-01) of the programme Investissements d'Avenir supervised
by the Agence Nationale de la Recherche (ANR).
LB acknowledges the European Research Council (ERC) project {\it
 Micromegas} and the ANR project {\it BlueEnergy}.
\end{acknowledgments}


%

\end{document}